\documentclass[jkps,article,fleqn,showpacs]{revtex4}
\usepackage{graphicx}
\usepackage{amssymb}
\usepackage{amsmath}
\usepackage{bm}
\usepackage{color}
\begin{document}

\begin{flushright}
{\tt arXiv:YYMM.NNNN}
\end{flushright}

\setcounter{page}{0}
\title{Multi-BPS D-vortices}
\author{Inyong \surname{Cho}}
\email{iycho@skku.edu}
\author{Taekyung \surname{Kim}}
\email{pojawd@skku.edu}
\author{Yoonbai \surname{Kim}}
\email{yoonbai@skku.edu}
\affiliation{BK21 Physics Research Division, Department of Physics,
and Institute of Basic Science,\\
Sungkyunkwan University, Suwon 440-746, Korea}

\date[]{}

\begin{abstract}
We investigate the BPS configuration of the multi D-vortices
produced from the D2${\bar {\rm D}}$2 system. Based on the DBI-type
action with a Gaussian-type runaway potential for a complex tachyon
field, the BPS limit is achieved when the tachyon profile is thin.
The solution states randomly-distributed $n$ static D-vortices with
zero interaction. With the obtained BPS configuration,
we derive the relativistic Lagrangian which describes the dynamics of
free massive D-vortices.
We also discuss the 90${}^{\circ}$ and 180${}^{\circ}$ scattering
of two identical D-vortices,
and present its implications on the reconnection in the dynamics of
cosmic superstrings.
\end{abstract}



\maketitle

\section{Introduction}
The Bogomolnyi-Prasad-Sommerfield (BPS) limit of solitons~\cite{Bogomolny:1975de}
has been attractive in its nature and tractability of dynamics.
The BPS limit is characterized by a minimum-energy configuration
of a static soliton which is a solution to the first-order
Bogomolnyi equation.
The dynamics of multi-BPS solitons is characterized by
no interaction. The BPS solitons do not exert any forces to
nearby solitons, so they move freely until they collide.

Recently D- and DF-strings have attracted attention as new candidates of
cosmic superstrings which produce
Y-junctions~\cite{Copeland:2003bj,Dvali:2003zj}.
Concerning string cosmology, such superstrings generated through
the decay of D3${\bar {\rm D}}$3 system are intriguing since the
separated D${\bar {\rm D}}$ provides
a natural setting for an inflationary epoch~\cite{Dvali:1998pa} and can be
employed in a string cosmological model with fluxes and moduli
stabilization~\cite{Kachru:2003sx}. Once D-strings are produced,
understanding their dynamics~\cite{Jackson:2004zg,Copeland:2006eh} is
important in order to see the distinctive
evolution between cosmic strings and cosmic superstrings.

In this work, we consider D-vortices which are produced in the
coincidence limit of D2${\bar {\rm D}}$2. (The model is equivalent
to parallel D-strings produced from D3${\bar {\rm D}}$3.) We
consider an effective field theory of a complex tachyon field
described by a Dirac-Born-Infeld (DBI) type action, which reflects
the instability of D${\bar {\rm D}}$
system~\cite{Sen:2003tm,Garousi:2004rd}. In the context of effective
field theory, D0-branes from D2${\bar {\rm D}}$2 have been obtained
as D-vortex configurations in
(1+2)-dimensions~\cite{Jones:2002si,Kim:2005tw,Cho:2005wp}.

In this article, we present a precise process obtaining the BPS
configuration for multi D-vortices and their moduli-space dynamics
based on Refs.~\cite{Kim:2006xi,Cho:2007jf}. The BPS limit is
possibly obtained with a Gaussian-type potential for the tachyon,
and the BPS configuration expresses infinitely thin D-vortices
randomly-distributed on a plane~\cite{Kim:2006xi}. We show also that
such a configuration reproduces the BPS sum rule and descent
relation correctly. With the obtained BPS solution, we derive a
relativistic Lagrangian which describes correctly the dynamics of
the multi BPS D-vortices~\cite{Cho:2007jf}. The BPS D-vortices
behave as free point-like particles with mass identified with the
D0-brane tension when they are separated.
We show that the scattering of two identical BPS D-vortices
exhibits the 90${}^{\circ}$ scattering and the 180${}^{\circ}$
(equivalently 0${}^{\circ}$) scattering,
which corresponds individually to the reconnection and the passing through
of D-strings when it is applied to the dynamics of cosmic superstrings.

The article is organized as following. In Section~II, we review the
Nielsen-Olesen vortex in the Abelian-Higgs model and present a
derivation of its BPS bound. In Section~III, we present precise BPS
conditions and obtain the BPS configuration of multi D-vortices. In
Section~IV, we derive the classical moduli-space dynamics of these
BPS D-vortices. We conclude in Section~V.

\section{Ablian-Higgs model and BPS limit}
In this section, we briefly review the Nielsen-Olesen vortex in
Abelian-Higgs system, and its BPS properties. The Abelian-Higgs
model is described by a complex scalar field $\phi$ and a U(1) gauge
field $A^\mu$. In flat $(1+2)$ dimensions, the action reads
\begin{equation}
S = \int dtd^2x \left[ -\frac{1}{2}\overline{D_\mu\phi}D^\mu\phi
-V(\phi) -\frac{1}{4}F_{\mu\nu}F^{\mu\nu} \right],
\label{Sno}
\end{equation}
where $F_{\mu\nu} = \partial_\mu A_\nu - \partial_\nu A_\mu$ is the
field-strength tensor of the gauge field and $D_\mu = \partial_\mu
-iA_\mu$ is the covariant derivative. The electric component of the
field strength is given by $E_i = F_{ti} = \partial_t A_i
-\partial_i A_t$ and the magnetic component is by $B = \partial_x
A_y - \partial_y A_x$.

\begin{figure}[t]
\includegraphics[width=110mm]{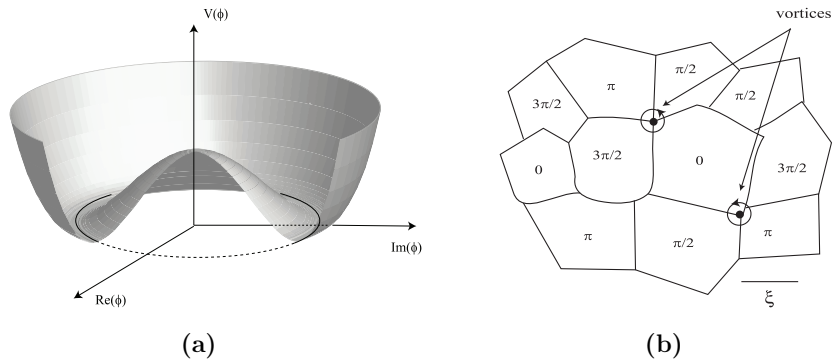}
\caption{(a) Plot of the unperturbed potential.
(b) Plot of mapping for a complex scalar field and vortex formation. }
\label{fig1}
\end{figure}

The Nielsen-Olesen vortex is produced during the phase transition
accompanied by spontaneous symmetry-breaking with a
temperature-dependent effective potential,
\begin{eqnarray}
V_{\rm eff}(\phi,T) = \frac{\alpha\lambda}{4} T^2 |\phi|^2 +V(\phi),\quad
V(\phi) = \frac{\lambda}{8} (|\phi|^2-\eta^2)^2,
\label{Veff}
\end{eqnarray}
where $\eta$ is the symmetry-breaking scale, and $\lambda$ is the
self-coupling constant, and $\alpha$ is a dimensionless constant.
The effective potential has a temperature-dependent effective mass
$m_{\rm eff}^2(T) = \lambda (\alpha T^2-\eta^2)/2$. At sufficiently
high temperatures, $T > T_c \equiv \eta/\sqrt{\alpha}$, $m_{\rm
eff}^2$ is positive, and the effective potential $V_{\rm eff}$ has
the minimum at $|\phi|=0$ which is the vacuum state. This vacuum is
invariant under the rotation in the field space, and the U(1)
symmetry is preserved. As the temperature drops, $m_{\rm eff}^2$
becomes negative, $V_{\rm eff}$ develops minima at $|\phi| = |m_{\rm
eff}|/\sqrt{\lambda}$, and the phase transition undergoes. When the
phase transition is completed at zero temperature, the system is
described by the unperturbed potential $V$. The vacuum manifold is
an ${\rm S}^1$ of radius $|\phi|=\eta$ in the phase field space as
shown in Fig.~\ref{fig1}-(a). Since the vacuum is superselected by a
point in this ${\rm S}^1$ and is not invariant under the rotation,
the continuous U(1) symmetry is broken.

As a result of phase transition, a vortex configuration can be formed.
The field configuration of the vortex has vanishing field $\phi$
(i.e., the symmetric state) at the center,
and the phase of $\phi$ increases by $2\pi n$ along a circle
enclosing the center counter-clockwise.
Here, $n$ is a nonzero integer which is interpreted as a winding number.
The negative value of $n$ represents an anti-vortex.

Such a nontrivial vortex configuration is guaranteed by the Kibble
mechanism~\cite{Kibble:1976sj}. As the phase transition proceeds,
the value of $\phi$ in the spatial points transits from the
symmetric phase (zero) to the broken symmetric one (nonzero). The
domains in the physical space are correlated only by a causal length
$\xi$ which is not larger than the horizon scale at that moment.
Therefore, each domain takes a different phase value in the vacuum
manifold. When the phases of domains are aligned in such a way to
wrap the circle of the vacuum manifold, it is irresistible to have a
boundary point of the domains, where the field vanishes to avoid a
singularity. Thereby, a vortex configuration is achieved. (See
Fig.~\ref{fig1}-(b).) Such a nontrivial configuration is called the
``vortex configuration" which exhibits $\phi = 0$ at the center and
$\phi = \eta$ in the asymptotic region. When the gauge field is
involved, it is called the Nielsen-Olesen
vortex~\cite{Nielsen:1973cs}. Once vortices are formed in space, it
is topologically stable since erasing the vortex configuration
requires the rearrangement of the field structure in entire infinite space.

Now, let us discuss the BPS picture of the Nielsen-Olesen
vortex~\cite{MS}. We assume a static-field configuration and the
Weyl gauge, $A_0 =0$. Then, the Bogomolnyi bound for the vortex is
saturated when $\lambda =1$. The energy-momentum tensor from the
action~\eqref{Sno} is
\begin{equation}
T_{\mu\nu}=\frac{1}{2}\left(\overline{D_\mu
\phi}D_\nu\phi+\overline{D_\nu \phi}D_\mu\phi\right)
+g^{\alpha\beta}F_{\mu\alpha}F_{\nu\beta}+g_{\mu\nu}{\cal
L}.\label{Ten}
\end{equation}
After rescaling
$\eta x^\mu \to x^\mu$, $\phi/\eta \to \phi$, and $A^\mu/\eta \to A^\mu$,
the Ginzburg-Landau energy evaluated from the
action~\eqref{Sno} becomes
\begin{eqnarray}
E &=& \int d^2x (-T^0_0) \\
&=& \frac{1}{2}\int d^2x \left[ B^2 + \overline{D_i\phi}D_i\phi
+\frac{1}{4} (\bar\phi\phi-1)^2 \right] \\
&=& \frac{1}{2}\int d^2x \left\{ \left[ B \pm \frac{1}{2}
\left(\bar\phi\phi -1 \right) \right]^2 + |(D_x \pm iD_y)\phi|^2
\pm B \mp \epsilon_{ij}\partial_iJ_j\right\}\label{GLE}\\
&\geq&\frac{1}{2}|\Phi|,\label{FL}
\end{eqnarray}
where $J_i=i\left({\bar\phi}D_i\phi-\phi{\overline{
D_i\phi}}\right)/2$ in Eq.~\eqref{GLE} is the conserved U(1)
current, the upper/lower sign is for vortex/anti-vortex in
Eq.~\eqref{GLE}, and $\Phi$ in Eq.~\eqref{FL} is magnetic flux,
\begin{equation}
\Phi\equiv\int d^2x B.
\end{equation}
The U(1) current vanishes at spatial infinity and does not
contribute to the energy after integration. For an object carrying
magnetic flux, the equality holds and the minimum-energy
configuration is achieved when the first two squared terms vanish,
which provides two Bogomolnyi equations,
\begin{equation}
(D_x\pm iD_y) \phi =0,\quad B \pm \frac{1}{2}\left(\bar\phi\phi -1
\right) = 0. \label{NOBog}
\end{equation}
Since the magnetic flux is quantized and uncharged by small
fluctuation, these two first-order equations should reproduce
second-order (Euler-Lagrange) field equations associated with the
original action~\eqref{Sno},
\begin{eqnarray}
D^iD_i\phi - \frac{1}{2}(\bar\phi\phi-1)\phi =0,\quad \epsilon_{ij}
\partial^j B +\frac{i}{2} (\bar\phi D_i\phi -\phi\overline{D_i\phi}
)=0. \label{NOEL}
\end{eqnarray}
Therefore, when the Nielsen-Olesen vortex (or anti-vortex) carrying
magnetic flux $\Phi=2n$ satisfies the Bogomolnyi
equations~\eqref{NOEL}, the Ginzburg-Landau energy in Eq.~\eqref{GLE}
saturates the Bogomolnyi bound in Eq.~\eqref{FL} and shows a so-called
BPS sum rule, \
\begin{eqnarray}
E = E_n = \pi|n|. \label{Eno}
\end{eqnarray}
Note that, in the Bogomolnyi limit of $\lambda=1$, the mass of gauge
field ($\eta$) and that of Higgs field ($\sqrt{\lambda}\eta$) become
the same, and this property is consistent with the observation that the
action~\eqref{Sno} is bosonic sector of the $N=2$ supersymmetric
extension of the Abelian-Higgs model~\cite{Di Vecchia:1977bs}.

Let us discuss the $n$ static
vortex configurations superimposed at the origin for arbitrary
$\lambda$ in what follows. From Eq.~\eqref{Ten}, its stress components
are rewritten as
\begin{eqnarray}
T^{x}_{\; x}&=&\frac{1}{4}\left[(\overline{D_{x}+ iD_{y})\phi}
(D_{x}- iD_{y})\phi +(\overline{D_{x}- iD_{y})\phi}
(D_{x}+ iD_{y})\phi \right] \nonumber\\
&&+\frac{1}{2}\left[B+\frac{\sqrt{\lambda}}{2}\left({\bar
\phi}\phi-\eta^2\right)\right]\left[B-\frac{\sqrt{\lambda}}{2}\left({\bar
\phi}\phi-\eta^2\right)\right],\label{TX}\\
T^{y}_{\; y}&=&-\frac{1}{4}\left[(\overline{D_{x}+ iD_{y})\phi}
(D_{x}- iD_{y})\phi +(\overline{D_{x}- iD_{y})\phi}
(D_{x}+ iD_{y})\phi \right] \nonumber\\
&&+\frac{1}{2}\left[B+\frac{\sqrt{\lambda}}{2}\left({\bar
\phi}\phi-\eta^2\right)\right]\left[B-\frac{\sqrt{\lambda}}{2}\left({\bar
\phi}\phi-\eta^2\right)\right],\label{TY}\\
T^{x}_{\;y}&=&\frac{i}{4}\left[(\overline{D_{x} \pm iD_{y})\phi}
(D_{x}\mp iD_{y})\phi -(\overline{D_{x} \mp iD_{y})\phi} (D_{x}\pm
iD_{y})\phi \right].\label{XY}
\end{eqnarray}
It is manifest that the solutions to the Bogomolnyi
equations~\eqref{NOBog} make these stress components vanish in the
BPS limit ($\lambda =1$). It means that the attractions and the repulsions are
completely canceled out everywhere for the static BPS Nielsen-Olesen
vortex configuration of arbitrary shape and winding number $n$.

For the $n$ superimposed vortices, the scalar and the gauge field
which satisfy the Euler-Lagrange
equations~\eqref{NOEL}, have  asymptotic expressions in cylindrical
coordinates $(\rho,\theta)$ as
\begin{eqnarray}\label{aex}
\phi (\rho) \sim 1- \frac{A_{\rm s}}{2\pi}K_0(\sqrt{\lambda}\rho),\quad
A_\theta (\rho) \sim n- \frac{A_{\rm m}}{2\pi} \rho K_1(\rho),
\end{eqnarray}
where $K_n$ is the modified Bessel function. The coefficients
$A_{\rm s}$ and $A_{\rm m}$ associated with the decay of scalar and
magnetic fields, are only determined numerically for nonBPS
vortices~\cite{de Vega:1976mi,Speight:1996px},
but, for BPS vortices, are predicted
analytically by the argument based on string duality~\cite{Tong:2002rq}.

As it was mentioned previously for the general BPS configuration,
a defining property of the BPS
configuration is implied in the interaction of vortices. Consider
two unit-winding vortices with a separation $s\gg 1$. The
interaction energy $E_{\rm int}(s)$ is the total energy $E$ minus
$2E_1$. There are two contributions to $E_{\rm int}$ coming from the
scalar field and the magnetic field. At large separations, the
interaction energy is evaluated from Eq.~\eqref{aex} and
obtained~\cite{Bettencourt:1994kf} by
\begin{eqnarray}
E_{\rm int}(s) =
-\frac{A_{\rm s}^2}{2\pi}K_0(\sqrt{\lambda}s)
+\frac{A_{\rm m}^2}{2\pi} K_0(s).
\end{eqnarray}
Naturally, the scalar field induces attraction while the magnetic
field does repulsion. For $\lambda=1$,
$A_{\rm s} = A_{\rm m}$~\cite{Speight:1996px,Tong:2002rq},
which means the interaction energy vanishes. It confirms that there
is no interaction between static vortices, when the BPS limit is saturated.
Since the interaction scale is of order $1/v$ and there does not exist
gapless radiation in the Abelian-Higgs model, the vortices move
freely until the separation becomes small enough ($1/v>1$).
For $\lambda <1$, the
scalar term dominates and the vortices attract. For $\lambda
>1$, the magnetic term dominates and the vortices repel.

As shown in the discussion based on the stress components~\eqref{TX}--\eqref{XY} of
the energy-momentum tensor, the interaction story in the
BPS limit can be extended to the $n$-vortex configurations and be
figured out also from the energy relation (the BPS sum
rule)~\eqref{Eno}. Consider an $n$-winding vortex. From the energy
relation, the interaction energy is given by $E_{\rm int} = E_n -
nE_1 =0$. This means that well-separated $n$ unit-winding vortices
do not interact to accumulate to a single $n$-winding vortex. In
addition, a single $n$-winding vortex does not dissociate to $n$
unit-winding vortices, either. Therefore, there will be no
interaction among vortices. For $\lambda <1$, $E_n < nE_1$, and the
vortices attract to accumulate. For $\lambda <1$, $E_n > nE_1$, and
the vortices repel to move apart.

The story so far is about the properties of BPS Nilesen-Olesen vortices.
The concept of the BPS limit in what follows would more precisely specified
by following three conditions.

\noindent {\bf (C1)} The static configurations are obtained from
first-order Bogomolnyi equations: These equations are usually
obtained by vanishing stress components of the energy-momentum
tensor, $T^i_j =0$, or by minimizing the energy.

\noindent {\bf (C2)} These BPS configurations should satisfy
Euler-Lagrange equations: For the case of static Nielsen-Olesen
vortex, it is automatic because the Bogomolnyi equations reproduce
Euler-Lagrange equations with $\lambda=1$.

\noindent {\bf (C3)} The BPS configurations provide the BPS-sum rule
as seen from Eq.~\eqref{Eno}: This states that the integrated energy
is given by a product of a certain topological charge, for example,
the winding number $n$ of vortex.

\section{D${\bar {\bf D}}$ System and BPS Limit of Multi-D-vortices}

The properties of D$(p-2)$ (or ${\bar {\rm D}}(p-2)$) produced from
the system of D$p{\bar {\rm D}}p$ in the coincidence limit is
described by an effective field theory of a complex tachyon field,
$T=\tau \exp(i\chi)$, and two Abelian gauge fields of
U(1)$\times$U(1) gauge symmetry, $A^{\mu}$ and $C^{\mu}$.  A
candidate is a DBI type action~\cite{Sen:2003tm,Garousi:2004rd}
\begin{equation}\label{ac}
S=-{\cal T}_{p}\int d^{p+1}x\,
V(\tau)\left[\,\sqrt{-\det(X^{+}_{\mu\nu})}
+\sqrt{-\det(X^{-}_{\mu\nu})}\,\,\right],
\end{equation}
where ${\cal T}_{p}$ is the tension of the D$p$-brane, $V(\tau)$ is
a runaway-type potential which is normalized as $V(\tau =0) =1$ and
vanishes as $V(\tau\to\infty) \to 0$, and
\begin{equation}\label{Xpm}
X^{\pm}_{\mu\nu}=g_{\mu\nu}+F_{\mu\nu}\pm C_{\mu\nu} +({\overline
{D_{\mu}T}}D_{\nu}T +{\overline {D_{\nu}T}}D_{\mu}T)/2
\end{equation}
with $ F_{\mu\nu}=\partial_{\mu}A_{\nu}-\partial_{\nu}A_{\mu},~
C_{\mu\nu}=\partial_{\mu}C_{\nu}-\partial_{\nu}C_{\mu},~ \mbox{and}~
D_{\mu}T=(\partial_{\mu} -2iC_{\mu})T. $

In this work, we shall consider only D${\bar {\rm D}}$ of $p=3$ in
type IIB string theory where both gauge fields $A^\mu$ and $C^\mu$
are turned off. Then, we obtain codimension-two object, D-strings
(D1-branes). Since we are interested in straight multi-D-strings
stretched parallel to an axis, parallel one-dimensional  D-strings
in three dimensions from D3${\bar {\rm D}}$3 can equivalently be
treated as point-like D-vortices in two dimensions. Considering a
static-tachyon configuration for multi-D-vortices in the
$(x,y)$-plane without quantized magnetic flux, we have
$F_{\mu\nu}=0$, $C_{\mu\nu}=0$, and
\begin{equation}\label{st}
T=T(x^{i}), \quad (i=1,2),
\end{equation}
and shall precisely derive the BPS limit of static multi-D-vortices
following the three BPS requirements presented in the previous
section.

\vspace{12pt} {\bf (C1) Bogomolnyi equation and its BPS solutions:}
Plugging the tachyon~\eqref{st} in the stress components of the
energy-momentum tensor leads to
\begin{equation}
T^{i}_{\;j}=-\frac{2{\cal T}_2
V}{\sqrt{1+S_{mm}-\frac{1}{2}A_{mn}^2}}
\left[\delta_{ij}-\left(S_{ij}-\delta_{ij}S_{kk}\right)+\left(A_{ik}A_{jk}
-\frac{\delta_{ij}}{2}A_{kl}^2\right) \right], \label{tij}
\end{equation}
where
\begin{equation}\label{sij}
S_{ij}\left(A_{ij}\right) =\frac{1}{2}\left(\partial_i{\overline{
T}}\partial_j T\pm \partial_j{\overline{ T}}\partial_i T\right).
\end{equation}
Considering the pressure difference and reshuffling the terms, we
obtain
\begin{equation}
T^{x}_{\; x}-T^{y}_{\; y} = \frac{{\cal T}_2
V}{\sqrt{1+S_{ii}-\frac{1}{2}A_{ij}^2}}
\left[(\overline{\partial_{x}T+ i\partial_{y}T}) (\partial_{x}T-
i\partial_{y}T) +(\overline{\partial_{x}T- i\partial_{y}T})
(\partial_{x}T+ i\partial_{y}T) \right]. \label{xxyy}
\end{equation}
It vanishes when the first-order Cauchy-Riemann equation,
\begin{equation}\label{BPS}
(\partial_{x}\pm i\partial_{y})T=0,\qquad (\,
\partial_x \ln \tau = \pm \partial_y \chi \;{\rm and} \;
\partial_y \ln \tau = \mp \partial_x \chi \,),
\end{equation}
is satisfied, and thus we may employ this as the Bogomolnyi equation.
Applying the Bogomolnyi equation to the off-diagonal stress
component $T^{x}_{\; y}$, we confirm that it vanishes
\begin{eqnarray}
T^{x}_{\; y}&=&i\frac{{\cal T}_2
V}{2\sqrt{1+S_{ii}-\frac{1}{2}A_{ij}^2}}
\left[(\overline{\partial_{x}T\pm i\partial_{y}T}) (\partial_{x}T\mp
i\partial_{y}T) -(\overline{\partial_{x}T \mp i\partial_{y}T})
(\partial_{x}T\pm i\partial_{y}T) \right]
\stackrel{(\ref{BPS})}{=}0.
\label{C1Txy}
\end{eqnarray}

We solve the Bogomolnyi equation~\eqref{BPS} and obtain static BPS
configuration of $n$ static D-vortices located randomly in the
$(x,y)$-plane. The ansatz of the tachyon field is
\begin{eqnarray}\label{ans}
\displaystyle{T = \tau(x,y) e^{i \sum_{p=1}^n\theta_p},}\qquad
\theta_p = \tan^{-1} \frac{y-y_p}{x-x_p},
\end{eqnarray}
where ${\bf x}_{p}= (x_{p},y_{p})$ $(p=1,2,...,n)$ denotes the
position of $p$-th D-vortex. Inserting the ansatz (\ref{ans}) into
the Bogomolnyi equation (\ref{BPS}), we obtain the solution for the
tachyon amplitude,
\begin{equation}\label{tau}
\tau(x,y)=\prod_{p=1}^{n}\tau_{{\rm BPS}}|{\bf x}-{\bf x}_p|,
\end{equation}
where $\tau_{{\rm BPS}}$ is a constant. Plugging the ansatz
(\ref{ans}) and solution (\ref{tau}) into the pressure components,
we obtain
\begin{eqnarray}
-T^{x}_{\; x}=-T^{y}_{\; y}=2{\cal T}_2 V.
\end{eqnarray}
In the zero-radius limit of D-vortices, $\tau_{{\rm BPS}}\rightarrow
\infty$, the potential $V$ vanishes everywhere except the centers of
vortices since $V$ is a runaway type. Therefore, the pressure
components vanish in this limit everywhere except the vortex
positions with $-T^{x}_{\; x}|_{{\bf x}={\bf x}_{p}}=-T^{y}_{\;
y}|_{{\bf x}={\bf x}_{p}} =2{\cal T}_2$. As a result, a candidate
solution to the Bogomolnyi equation~\eqref{BPS} is obtained by
taking the thin limit,
\begin{eqnarray}
\tau(x,y)=\lim_{\tau_{{\rm
BPS}}\to\infty} \prod_{p=1}^{n}\tau_{{\rm BPS}}|{\bf x}-{\bf x}_p|
=\left\{%
\begin{array}{ll}
0, & \hbox{at each ${{\bf x}}_{p}$}\,, \\
\infty, & \hbox{elsewhere}. \\
\end{array}%
\right.
\label{BGsol}
\end{eqnarray}

\vspace{12pt} {\bf (C2) Euler-Lagrange equation:} For the
nontrivial-tachyon configuration, the Euler-Lagrange equation is
equivalent to the conservation of the energy-momentum tensor. Since
$T^x_{\; y} =0$ completely as seen in Eq.~\eqref{C1Txy}, the
energy-momentum conservation $\partial_i T^{i}_{\; j}=0$ reduces to
$\partial_x T^x_{\; x} =0$ and $\partial_y T^y_{\; y} =0$.
As it was discussed previously in ({\bf  C1}), $T^{x}_{\; x}$ and $T^{y}_{\;
y}$ are finite at the positions of D-vortices and vanish elsewhere
in the $\tau_{\rm BPS} \to \infty$ limit. Therefore, $x$- and
$y$-component of the conservation of the energy-momentum tensor hold
when the derivatives are considered as weak
derivatives~\cite{Evans,Go:2007fv}.

\vspace{12pt} {\bf (C3) BPS sum rule and descent relation:} The BPS
solution should reproduce a correct BPS sum rule
from the integrated-energy expression. Here we show that the solution \eqref{BGsol}
possibly reproduces the BPS sum rule under a Gaussian-type
potential,
\begin{equation}\label{bsft}
V(\tau)=\exp\left(-\frac{\tau^{2}}{\pi R^{2}}\right).
\end{equation}
The computation of Hamiltonian for $n$ randomly-located D-vortices
\eqref{ans}-\eqref{tau} reproduces the BPS sum rule,
\begin{eqnarray}
E = {\cal T}_{0}|n|= \int d^2 x \;{\cal H}_{\rm BPS}&=&2{\cal T}_2 \int
d^2x \; \lim_{\tau_{{\rm BPS}}\rightarrow \infty} V\left(\tau
\right) (1+S_{xx})
\label{ham}\\
&=& 2{\cal T}_2 \int d^2x \; \lim_{\tau_{{\rm BPS}}\rightarrow
\infty} V\left(\tau \right) S_{xx}
\label{sxc}\\
&=&2\pi^{2}R^{2}{\cal T}_{2}|n|,
\label{dsr}
\end{eqnarray}
where ${\cal T}_{0}$ denotes the mass of unit D-vortex. The last
line (\ref{dsr}) means that the descent relation for codimension-two
BPS branes, ${\cal T}_{0}=2\pi^{2}R^{2}{\cal T}_{2}$, is correctly
obtained, and the Hamiltonian density for BPS configurations is
expressed in terms of a sum of $\delta$-functions,
\begin{eqnarray}
{\cal H}_{{\rm BPS}} &=& 2{\cal T}_{2}\lim_{\tau_{{\rm
BPS}}\rightarrow\infty}V(\tau)(1+S_{xx})
\label{Hbps} \\
&=& {\cal T}_{0} \sum_{p=1}^{n}\delta^{(2)}({\bf x}-{\bf x}_{p}(t)).
\label{De}
\end{eqnarray}
Note that, for $n$ superimposed D-vortices with rotational symmetry,
the integration in Eq.~(\ref{ham}) yields the correct descent
relation without taking the infinite $\tau_{{\rm BPS}}$ limit (or
the BPS limit).

From now on, we present the integration process in
Eqs.~\eqref{ham}-\eqref{dsr} more precisely. We compute the value of
energy for $n$ randomly-located BPS D-vortices of which tachyon
profile is given by Eqs.~\eqref{ans} and \eqref{BGsol}. With  the
potential~\eqref{bsft},  the integrated energy \eqref{ham}
becomes
\begin{eqnarray}
E= 2{\cal T}_2  \lim_{\tau_{\rm BPS}\to \infty} \int d^2 \tilde{x}
\; \exp\left(-\frac{1}{\pi R^2}\displaystyle{\prod_{p=1}^{n}}
|\tilde{{\bf x}} -\tilde{{\bf x}}_p|^{2} \right) \left[\frac{1}{
\tau_{\rm BPS}^2} + \left( \sum_{q,r=1}^n \frac{\cos\theta_{qr}}
{|\tilde{{\bf x}} -\tilde{{\bf x}}_q||\tilde{{\bf x}} -\tilde{{\bf
x}}_r|} \right) \displaystyle{\prod_{p=1}^{n}} |\tilde{{\bf x}}
-\tilde{{\bf x}}_p|^{2}\right],
\end{eqnarray}
where $\theta_{qr}$ is the angle between two vectors,
${\bf x}-{\bf x}_q$ and ${\bf x}-{\bf x}_r$, and
$|\tilde{{\bf x}} -\tilde{{\bf x}}_p| = \tau_{\rm BPS} |{\bf x}-{\bf x}_p|$
is a dimensionless variable. The first term with
$1/\tau_{\rm BPS}^2$ in the square-bracket becomes subdominant  for
sufficiently large $\tau_{\rm BPS}$ and vanishes in the limit of
$\tau_{\rm BPS}\rightarrow \infty$ . The contribution to integration
comes from the second term which we will separate into $n$ diagonal
and $n(n-1)$ off-diagonal components as
\begin{eqnarray}
\hspace{-8mm}E= 2{\cal T}_2 \lim_{\tau_{\rm BPS}\to \infty} \int d^2
\tilde{x} \; \exp\left(-\frac{1}{\pi
R^2}\displaystyle{\prod_{p=1}^{n}} |\tilde{{\bf x}} -\tilde{{\bf
x}}_p|^{2} \right) \left( \sum_{q=1}^n\frac{1}{|\tilde{{\bf x}}
-\tilde{{\bf x}}_q|^2} + \sum_{q=1}^n \sum_{r=1 (\neq q)}^n
\frac{\cos \theta_{qr}}{|\tilde{{\bf x}} -\tilde{{\bf
x}}_q||\tilde{{\bf x}} -\tilde{{\bf x}}_r|} \right)
\displaystyle{\prod_{p=1}^{n}} |\tilde{{\bf x}} -\tilde{{\bf
x}}_p|^{2}.
\label{energy}
\end{eqnarray}
Let us consider first the off-diagonal terms ($q\neq r$) assuming
that the distance between any pair of two vortices is always much
larger than $1/\tau_{\rm BPS}$. The integration of the off-diagonal
part is symmetric under the exchange $q \leftrightarrow r$, so we
may consider only for a given vortex located at ${\bf x}_q$. (We
will use the dimensional variable in the following analyses.) Off
the vortex locations, ${\bf x}\ne {\bf x}_{p}~(p=1,2,...,n)$, the
integrand of every term vanishes by taking the limit since
\begin{eqnarray}
\lim_{\tau_{{\rm BPS}\rightarrow \infty}}\exp(-{A}\tau_{{\rm
BPS}}^{\;\; 2n})B \tau_{{\rm BPS}}^{\;\; 2(n-1)}\rightarrow 0,
\end{eqnarray}
for any nonvanishing finite $B$ and positive $A$. At the vortex
locations, ${\bf x}={\bf x}_p$, $\tau =0$ and $V(\tau)=1$ from
Eqs.~\eqref{tau} and~\eqref{bsft}. In addition, we have
\begin{eqnarray}
S_{xx}\propto
\left\{%
\begin{array}{ll}
\left. |{\bf x}-{\bf x}_{p}|\right|_{{\bf x}={\bf x}_{p}} \rightarrow 0,
& \hbox{for $p=q$}\,, \\
\left. |{\bf x}-{\bf x}_{p}|^{2}\right|_{{\bf x}={\bf x}_{p}} \rightarrow 0,
& \hbox{for $p\neq q$}. \\
\end{array}%
\right.
\end{eqnarray}
Therefore, the off-diagonal ($q\ne r$) terms
do not contribute to the integrated BPS sum rule (\ref{dsr}) as
far as every pair of two D-vortices is separated with a distance
much larger than $1/\tau_{{\rm BPS}}$.
If we finally take $\tau_{{\rm BPS}}\rightarrow \infty$,
the integration values of all the off-diagonal terms
become zero irrespective of their separations.

Now let us consider the diagonal terms ($q=r$).
For this case, $S_{xx}$ can be written as
\begin{eqnarray}
S_{xx} = \sum_{q=1}^n\frac{1}{|\tilde{{\bf x}} -\tilde{{\bf
x}}_q|^2} \displaystyle{\prod_{p=1}^{n}} |\tilde{{\bf x}}
-\tilde{{\bf x}}_p|^{2} =\sum_{q=1}^n\displaystyle{\prod_{p=1 (\neq
q)}^{n}} |\tilde{{\bf x}} -\tilde{{\bf x}}_p|^{2} \equiv
\sum_{q=1}^n S_{xx}^q.
\end{eqnarray}
Off the vortex locations, ${\bf x}\ne {\bf x}_{p}$,
the same argument as the off-diagonal case
is applied and the integrand vanishes.
The source of nonvanishing integration
comes from the region near the vortex positions.
It is easy to see from the above equation that
for $p\neq q$, $S_{xx}^q$ vanishes at ${\bf x}={\bf x}_{p}$
while $V(\tau)\to 1$.
This does not contribute to integration.
At ${\bf x}={\bf x}_{q}$, however,
$S_{xx}^q$ diverges in the limit of $\tau_{\rm BPS}\to\infty$,
\begin{equation}\label{apr}
S_{xx}^q =
\left\{
\begin{array}{ll}
\left. \cdots |{\bf x}-{\bf x}_{p}|^{2}\cdots
\right|_{{\bf x}={\bf x}_{p}}\rightarrow 0,
& \mbox{for~} p~(\ne q), \\
\displaystyle{ \lim_{\tau_{{\rm BPS}}\rightarrow \infty}
\prod_{p=1 (\ne q)}^{n} (\tau_{{\rm BPS}}|{\bf x}_{q}-{\bf
x}_{p}|)^{2} \rightarrow \infty }, & \mbox{for~} q.
\end{array}
\right.
\end{equation}

We can perform now the energy integration \eqref{energy} with a
sufficient accuracy since we first take the $\tau_{{\rm
BPS}}\to\infty$ limit  for the BPS D-vortices. Using
Eq.~\eqref{apr}, in the limit of $\tau_{{\rm BPS}}\rightarrow
\infty$, the energy is rewritten as
\begin{eqnarray}
E= 2{\cal T}_2
\lim_{\tau_{\rm BPS}\to \infty} \sum_{q=1}^n \int d^2 \tilde{x} \;
\exp\left[-\frac{1}{\pi R^2}\left( \displaystyle{\prod_{p=1 (\neq
q)}^{n}} |\tilde{{\bf x}}_q -\tilde{{\bf x}}_p|^{2}\right)
|\tilde{{\bf x}} -\tilde{{\bf x}}_q|^{2} \right] \left(
\displaystyle{\prod_{p=1 (\neq q)}^{n}} |\tilde{{\bf x}}_q
-\tilde{{\bf x}}_p|^{2}\right).
\label{energy2}
\end{eqnarray}
Since the integrand is a Gaussian type, the integration is easily
performed in a closed form and its value is independent of
$\tau_{\rm BPS}$. In synthesis, the integration gives the BPS sum
rule obtained in Eq.~\eqref{dsr}. The analyses here are applied also
to the case of that arbitrary number of D-vortices are superimposed.
In summary, we have shown that the BPS sum rule and the correct
descent relation are reproduced by the solution \eqref{BGsol} of the
Bogomolnyi equation with the Gaussian-type potential~\eqref{bsft}.

\vspace{12pt} We have shown that the static multi-D-vortices in the
zero-radius limit fulfill the three BPS requirements. First, the
pressures vanish everywhere, $T^{x}_{\; x}=T^{y}_{\; y}=0$, except
the positions of D-vortices, and the off-diagonal stress vanishes
completely, $T^{x}_{\; y}=0$. This shows that separated D-vortices
are necessarily noninteracting. Second, the nontrivial D-vortex
configuration given by the solution to the first-order
Cauchy-Riemann equation also satisfies the conservation of the
energy-momentum tensor when the involved derivatives were weak
derivatives. This was equivalent to a satisfaction of the
Euler-Lagrange equation. Third, with a Gaussian-type tachyon
potential, the integrated energy of static $n$ D-vortices shows that
the BPS sum rule and the descent relation for codimension-two BPS
branes are correctly reproduced. Therefore, the fulfillment of these
necessary requirements suggests that a BPS limit of multi-D-vortices
from D3${\bar {\rm D}}$3 is achieved, and that the Cauchy-Riemann
equation can be identified with the first-order Bogomolnyi equation.
Since supersymmetry does not exist in the D3${\bar {\rm D}}$3
system, the derivation of BPS bound is lacked differently from the
usual BPS vortices in Abelian- Higgs model. In this sense, the BPS
properties of these multi- D-vortices need further study.

\section{Classical Dynamics of multi-BPS D-vortices}

In this section, we discuss the moduli-space dynamics of $n$ BPS
D-vortices. The BPS codimension-two D-vortices (D0-branes) at hand
are infinitely thin (zero-radius), so their classical dynamics may
be approximated by the motion of $n$-point particles in two
dimensions. The BPS nature predicts a free motion of separated
particles. The interaction occurs only in the range of collisions,
which is the coalescence limit for our thin BPS D-vortices.

\begin{figure}[t]
\includegraphics[width=40mm]{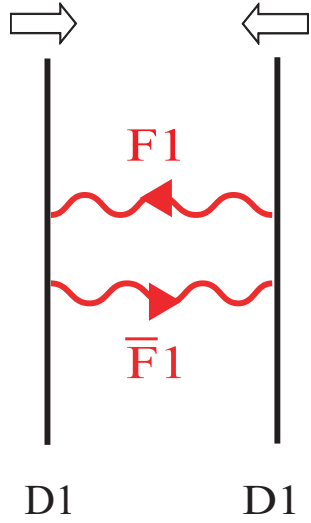}
\caption{Pair production of fundamental strings for colliding BPS
D-strings }\label{fig2}
\end{figure}

If we consider the classical dynamics of the thin BPS D-vortices in
the $(x,y)$-plane, the pair production of fundamental string (F1)
and anti-fundamental string (${\overline{\rm F}1}$) as shown in
Fig.~\ref{fig2} is neglected for colliding BPS D-strings. Although
the zero modes are not identified completely different from
multi-BPS Nielsen-Olesen vortices~\cite{Weinberg:1979er}, we assume
that the time-dependence of the field naturally appears only in the
vortex positions,
\begin{equation}\label{tde}
{\bf x}_{p}(t)=(x_{p}(t),y_{p}(t)).
\end{equation}
We begin with the tachyon-field configuration in
Eqs.~\eqref{ans}--\eqref{tau},
\begin{eqnarray}
\tau=\lim_{\tau_{{\rm BPS}}\rightarrow \infty}
\prod_{p=1}^{n}\tau_{{\rm BPS}}|{\bf x}-{\bf x}_p(t)|,\quad \chi =
\pm \sum_{p=1}^n \tan^{-1} \frac{y-y_p(t)}{x -x_p(t)}, \label{tauxi}
\end{eqnarray}
and the moduli-space dynamics can be described by the time evolution
of the vortex positions. Different from the moduli-space dynamics of
BPS Nielsen-Olesen vortices with finite
thickness~\cite{Shellard:1988zx}, there is a possibility to derive
the classical Lagrangian of the $n$ point BPS D-vortices in an exact
form from the action \eqref{ac} by integrating the Lagrangian
density over the space,
\begin{eqnarray}
{}\hspace{-4mm} L({\bf x}_{p}(t),{\dot {\bf x}}_{p}(t)) &\equiv&\int
d^{2}x\lim_{\tau_{{\rm BPS}}\rightarrow \infty}\, {\cal
L}(\tau,\chi,\partial_{\mu}\tau,\partial_{\mu}\chi)
\label{fla}\\
&=&-\int d^2 x \lim_{\tau_{{\rm
BPS}}\rightarrow \infty} {\cal H}_{\rm
BPS}\sqrt{1+\frac{\tau^2\left[-(\partial_{0}\ln
\tau)^2-(\partial_{0}\chi)^2-(\partial_{0}\tau\partial_{i}\chi
-\partial_{i}\tau\partial_{0}\chi)^{2}\right]}{\left(1+S_{xx}\right)^2}}\,.
\label{lag2}
\end{eqnarray}
The delta-function property \eqref{De} results from the energy
integration performed in the previous section and will be used in
what follows. Using the tachyon profiles in Eq.~\eqref{tauxi}, the
derivatives are obtained as
\begin{eqnarray}
\partial_{0}\ln\tau&=&-\sum_{p=1}^{n}{\dot{\bf x}}_{p}(t)\cdot
\frac{\partial \tau}{\partial({\bf x}-{\bf x}_{p}(t))}
= -\sum_{p=1}^{n}\frac{{\dot{\bf x}}_{p}(t)\cdot({\bf x}-{\bf
x}_{p}(t))}{\left|{\bf x}-{\bf x}_{p}(t)\right|^2},\quad
\partial_{i}\ln\tau = \sum_{p=1}^{n}\frac{({\bf
x}-{\bf x}_{p}(t))^i}{\left|{\bf x}-{\bf x}_{p}(t)\right|^2},\label{ita}\\
\partial_{0}\chi&=&-\sum_{p=1}^{n}{\dot{\bf x}}_{p}(t)\cdot
\frac{\partial \chi}{\partial({\bf x}-{\bf x}_{p}(t))}
=\sum_{p=1}^{n}\frac{\epsilon_{ij}{\dot{\bf x}_{p}^{i}}\,({\bf
x}-{\bf x}_{p}(t))^j}{\left|{\bf x}-{\bf x}_{p}(t)\right|^2},\quad
\partial_{i}\chi = -\sum_{p=1}^{n}\frac{\epsilon_{ij}({\bf
x}-{\bf x}_{p}(t))^j}{\left|{\bf x}-{\bf
x}_{p}(t)\right|^2}.\label{ich}
\end{eqnarray}
Plugging these derivatives into Eq.~\eqref{lag2} and using the
expression of the Hamiltonian density~\eqref{Hbps} in terms of the
tachyon profiles, we have
\begin{eqnarray}
L({\bf x}_{p}(t),{\dot {\bf x}}_{p}(t)) &=&
-{\cal T}_{0}
\int d^{2} x \sum_{s_{1}=1}^{n}
\delta^{(2)}({\bf x}-{\bf x}_{s_{1}}) \nonumber\\
&\times & \lim_{\tau_{{\rm BPS}}\rightarrow\infty}
\sqrt{1-\frac{ \displaystyle{
\left(\prod_{s_{2}=1}^{n}\tau_{{\rm BPS}} |{\bf x}-{\bf x}_{s_{2}}|
\right)^{2}\sum_{p,q=1}^{n}\frac{\cos \theta_{pq}}{ \tau_{{\rm
BPS}}|{\bf x}-{\bf x}_{p}| \tau_{{\rm BPS}}|{\bf x}-{\bf x}_{q}|}}
}{ \displaystyle{ \frac{1}{\tau_{{\rm
BPS}}^{2}}+\left(\prod_{s_{3}=1}^{n}\tau_{{\rm BPS}} |{\bf x}-{\bf
x}_{s_{3}}| \right)^{2}\sum_{s_{4},s_{5}=1}^{n}\frac{\cos
\theta_{s_{4}s_{5}}}{ \tau_{{\rm BPS}}|{\bf x}-{\bf x}_{s_{4}}|
\tau_{{\rm BPS}}|{\bf x}-{\bf x}_{s_{5}}|} } }{\,\dot {\bf x}}_{p}
\cdot {\dot {\bf x}}_{q} }\; .\label{tla}
\end{eqnarray}
Performing the delta-function integration and taking the $\tau_{{\rm
BPS}}\rightarrow\infty$ limit, we obtain the expected result
\begin{eqnarray}
L({\bf x}_{p}(t),{\dot {\bf x}}_{p}(t))
=-{\cal T}_{0}\sum_{p=1}^{n}\sqrt{1-{\dot {\bf x}}_{p}^{2}}\,.
\label{flg}
\end{eqnarray}
This Lagrangian describes $n$ relativistic free particles of mass
${\cal T}_{0}$ in the speed limit $|{\dot {\bf x}}_{p}|\le 1$ and
outside the range of mutual interaction. It correctly reflects the
character of point-like classical BPS D-vortices of which actual
dynamics is governed by the relativistic field equation of a complex
tachyon $T(t,{\bf x})$ and the gauge fields of U(1)$\times$U(1)
symmetry. In addition, since the size of the BPS D-vortex is zero,
the free Lagrangian description~\eqref{flg} is valid for any case of
small separation between two D-vortices. The relativistic Lagrangian~(\ref{fla})
of multi-BPS objects has never been derived through
systematic studies of moduli-space dynamics.

\begin{figure}[!t]
\begin{center}
\scalebox{0.6}[0.6]{\includegraphics{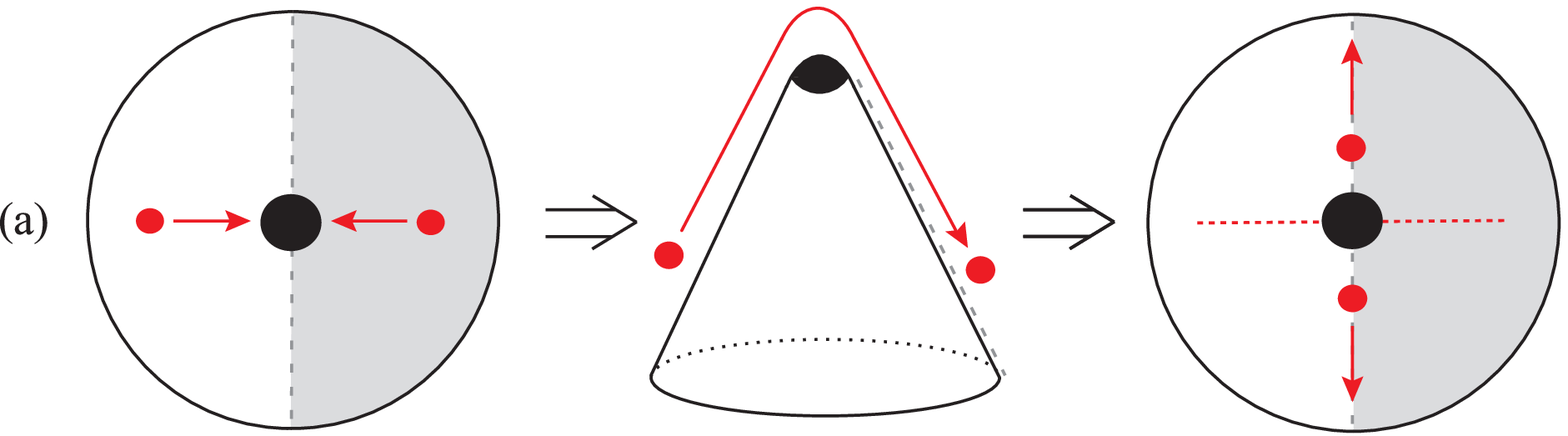}}\\
\scalebox{0.6}[0.6]{\includegraphics{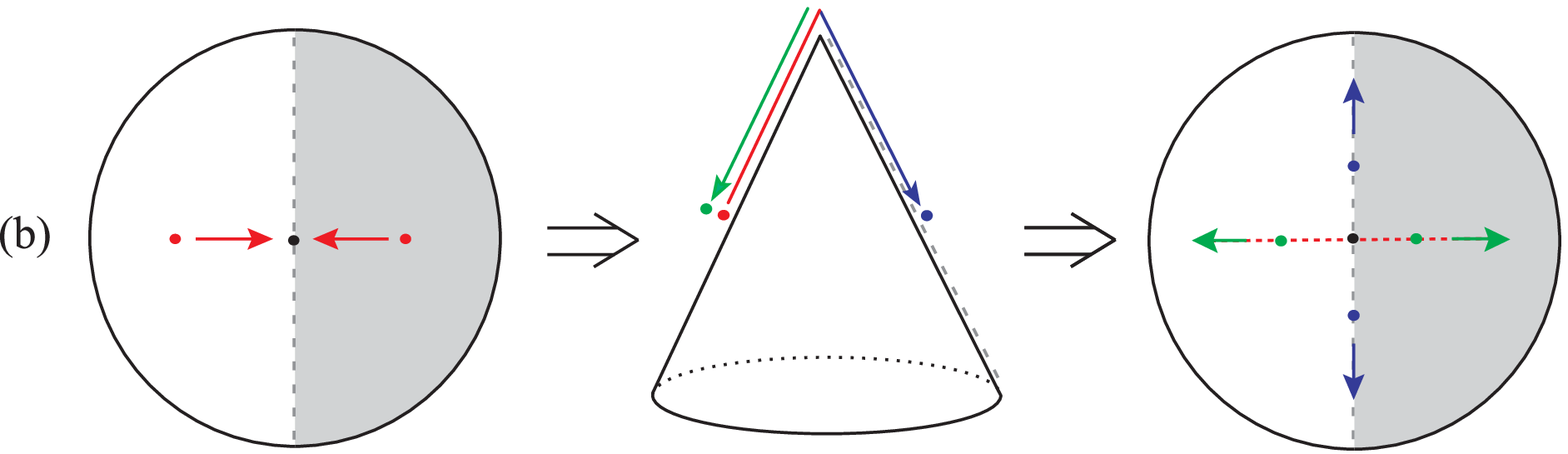}}
\par
\vskip-2.0cm{}
\end{center}
\caption{\small Scattering picture of (a) BPS Nielsen-Olesen vortices
and (b) two identical BPS D-vortices.
Since the two vortices are identical, a vortex sees only the half
space (the shaded region is equivalent to the unshaded region). In
addition, there is a ${\rm Z}_2$ symmetry between the upper and the
lower quadrant. After semi-diametric dashed lines are identified,
the moduli space for a vortex becomes a cone. For the finite-size
Nielsen-Olesen vortices, the moduli space is a stubbed cone, and
there is only the 90${}^{\circ}$ scattering. For the zero-size
D-vortices, the moduli space is a sharp cone. Considering the
symmetry the scattering has two possibilities, 90${}^{\circ}$
scattering and 0${}^{\circ}$ (equivalently 180${}^{\circ}$)
scattering.}   \label{fig3}
\end{figure}

\begin{figure}[!h]
\begin{center}
\includegraphics[width=165mm]{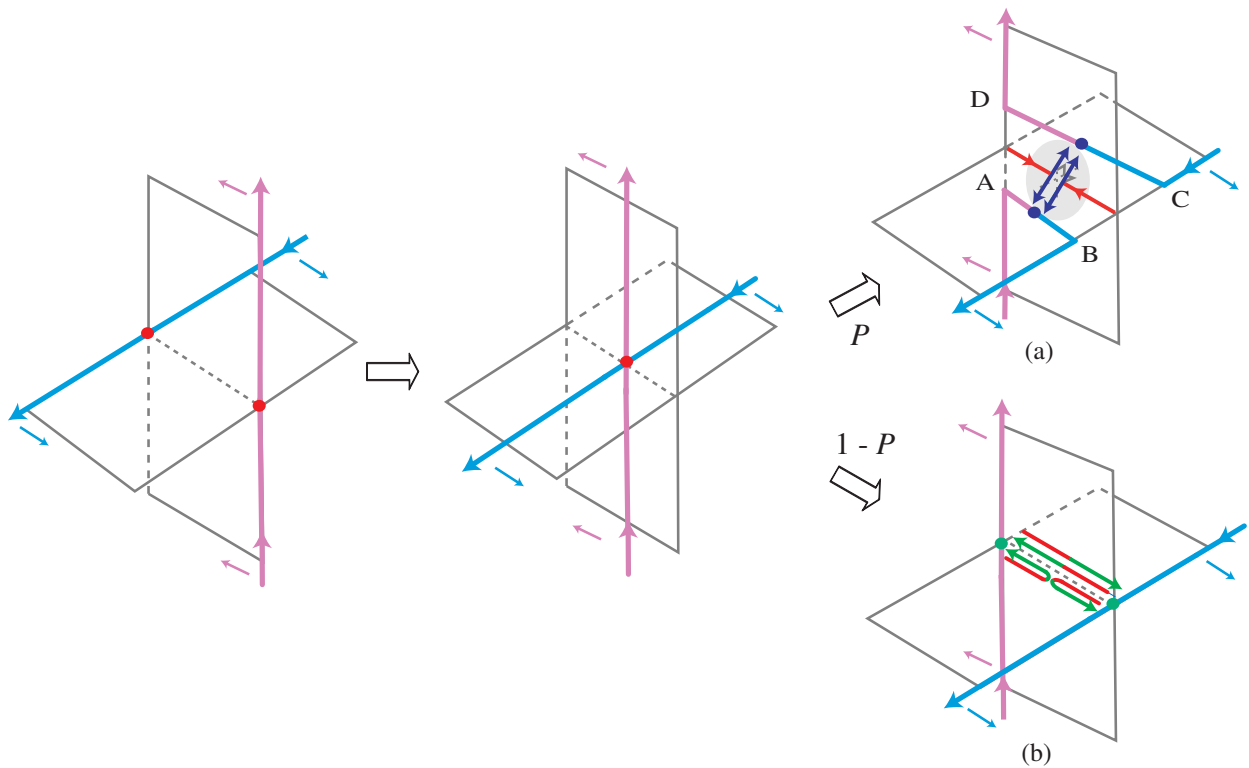}
\end{center}
\caption{Scattering picture of two identical BPS D-strings for (a)
reconnection (corresponding to 90${}^{\circ}$ scattering of
D-vortices) and (b) passing through (corresponding to
$0^{\circ}$/180${}^{\circ}$ scattering of D-vortices). In the
figures, the arrow on the string represents the signature of
vorticity, and the arrows perpendicular to the string represent the
direction of motion. In the two figures on the right, the arrows for
D-vortex scattering in Fig.~3-(b) are presented. In all of the four
figures, the scattering points and their trajectories before
collision are marked as red for both strings since they are
identical. In (a), the trajectories of D-vortices after
$90^{\circ}$ scattering are given by the blue lines.
The reconnection of D-strings is
possible only for pairs of ${\rm A}$-${\rm B}$ and ${\rm C}$-${\rm
D}$ since the other pairs of reconnection is not allowed considering
the vorticity. In (b), the trajectories after $0^{\circ}$ and
$180^{\circ}$ scattering are given by the green lines, both of which lead to the
passing through of two identical BPS D-strings.} \label{fig4}
\end{figure}

The methods of the moduli-space dynamics of multi-BPS vortices with
the canonical form of kinetic term assume a slow motion of BPS
solitons, and then read the metric of moduli
space~\cite{Shellard:1988zx,MS}. Therefore, its relativistic regime
is supplemented only by numerical analysis which solves field
equations directly. The relativistic Lagrangian~\eqref{fla} obtained
here is free from perturbative open string degrees due to the decay
of unstable D${\bar {\rm D}}$. It means that the classical dynamics
of BPS D-vortices with nonzero separation can be safely described by
Eq.~\eqref{fla}, and thus by Eq.~\eqref{flg}, and is consistent with
the numerical analysis dealing with time-dependent field equations.
One may also ask whether or not this relativistic Lagrangian of free
particles is a consequence of the DBI type action. The specific
question is how much the square-root form of the DBI
action~\eqref{ac} plays a role in derivation. Although we do not
have any other example to compare, the Lagrangian~\eqref{fla}
supports the validity of the DBI-type action~\eqref{ac} as a
tree-level Lagrangian.

The results of D-vortices in our model describe the parallel
D-strings obtained from the coincidence limit of D3${\bar {\rm
D}}$3. Finally we consider the collision of two identical BPS
D-vortices in Ref.~\cite{Cho:2007jf}. The classical dynamics of
cosmic D-strings can be read off from our results. For a collision
of two Nielson-Olesen vortices, one vortex sees only a half of the
space since the two vortices are identical. The resulting moduli
space becomes, therefore, a cone with a stubbed apex due to the
finite size of the vortex. The geodesic motion in this moduli space
is overcoming the apex straightly, which corresponds to the
right-angle scattering in the physical space~\cite{Shellard:1988zx}.
(See the red lines in Fig.~\ref{fig3}-(a).) Applied to crossing
cosmic strings, this right-angle scattering results in the
reconnection of the strings after scattering. (See the
Fig.~\ref{fig4}-(a).) This is the only possible motion for
Nielsen-Olesen cosmic strings, and the reconnection probability is
always one, $P=1$~\cite{Shellard:1988ki}.

For the identical BPS D-vortices, in addition to the right-angle
scattering (see the blue lines in Fig.~\ref{fig3}-(b)), there also
exist $0^{\circ}$ (passing-through) and $180^{\circ}$ scattering
motions (see the green lines in Fig.~\ref{fig3}-(b)) which are
indistinguishable. Since the BPS D-vortices are thin, the moduli
geometry is a cone with a sharp singular apex. The motion at the tip
of the cone is then unpredictable, and what we can apply is only the
symmetry argument; the vortex motion is either overcoming the tip or
bouncing back there.
The bouncing-back motion of D-vortices in the moduli space
corresponds to the $0^{\circ}$ or $180^{\circ}$ scattering
of two identical cosmic D-strings crossing in the physical space.
(See Fig.~\ref{fig4}-(b).) This affects the reconnection probability
of cosmic superstrings and one has $P<1$. This classical result with
the help of a quantum concept of identical particles predicts only
the possible scattering patterns.
The probability $P$ is determined by taking into account
the quantum correction as shown in Fig.~\ref{fig2}.

\section{Conclusions}
In this work, we obtained the BPS solution to the multi D-vortices
produced in the coincidence limit of D2${\bar {\rm D}}$2. The model
is described by a DBI-type action with a complex tachyon field. With
a gaussian type-potential, we showed that the BPS limit is achieved
by an infinitely thin tachyon profile. The obtained BPS
configuration correctly reproduces the BPS sum rule and descent
relation which result in the identification of BPS D-vortices as BPS
D-branes of codimension two.

For multi BPS D-vortices, it is expected that the thin vortices are
point-like particles moving freely when they are separated. Assuming
that the time-dependence of the system is encoded only in the vortex
positions in dynamics, we integrated the Lagrangian of the DBI
action to obtain a simple Lagrangian $L^{(n)}({\bf x}_{p},{\dot {\bf
x}}_{p}) = -{\cal T}_{0}\sum_{p=1}^{n}\sqrt{1-{\dot {\bf
x}}_{p}^{2}}$ which describes $n$ free relativistic point particles
with the mass given by the D0-brane tension.

While the scattering of Nielsen-Olesen vortices exhibits only the
90${}^{\circ}$ scattering, the scattering of BPS D-vortices exhibits
also the $0^{\circ}$/180${}^{\circ}$ scattering. When this
scattering picture is applied to the interaction of cosmic
superstrings, it implies that two identical BPS D-strings can pass
through each other ($0^{\circ}$/180${}^{\circ}$ scattering) as well
as can reconnect (90${}^{\circ}$ scattering). This is one of the key
differences of cosmic superstrings from usual cosmic strings.

\begin{acknowledgments}
This work is the result of research activities (Astrophysical
Research Center for the Structure and Evolution of the Cosmos
(ARCSEC)) and grant No.\ R01-2006-000-10965-0 from the Basic
Research Program supported by KOSEF.
I.C. and Y.K. are grateful to the organizers and the Asia Pacific Center
for Theoretical Physics
for hospitalities and supports for the School on Black Hole Astrophysics 2008.
\end{acknowledgments}

\end{document}